# Bioinspired Associative Memory System Based on Enzymatic Cascades

Kevin MacVittie,[a] Jan Halámek,[a] Vladimir Privman,[b] and Evgeny Katz[a]*

**A biomolecular system representing the first realization of associative memory based on enzymatic reactions *in vitro* has been designed. The system demonstrated "training" and "forgetting" features characteristic of memory in biological systems, but presently realized in simple biocatalytic cascades.**

Among ideas and developments aiming at new bio-inspired functionalities for signal and information processing, devices involving memory, particularly those based on molecular/supramolecular[1] and nanoscale systems,[2] have drawn a significant recent interest. Recent research in electronics and materials science resulted in novel elements with memory functions including memristors[3,4] and other designs.[5] Despite the fact that many of such functionalities are modeled after mechanisms common in nature, there have been few realizations involving biomolecular processes.[6-8] Information processing with biomolecules has been an active field[9] offering promise and recent successes in biocompatible interfacing,[10] multi-input sensing,[11] and generally new supplementary capabilities to electronic devices.[12] However, this research has been primarily concerned with the design and study of binary gates[13,14] and small networks[15,16] carrying out digital processing steps. These involved standard gates, AND, OR, XOR, etc., which can be carried out by DNA/RNA,[17-19] proteins/enzymes[14,20] and cells.[21,22] Few-step enzymatic-process networks have been devised for specific sensing[11] and diagnostic applications,[23,24] and aspects of noise-handling properties of such "biocomputing" networks and their scalability have been explored,[25] as were, to some extent, aspects of their control by and/or interfacing with electrodes and physical signals.[12]

Development of processes with functionalities such as memory by mechanisms suggested by nature but based on a simple set of biomolecular processes rather than a full complexity of living systems, is of interest for many reasons. New paradigms for information processing might be possible beyond the presently used approaches of analog and digital electronics. Presently there is no complete understanding of how natural systems manage complex, large-scale information processing.[26] Therefore, researching model biomolecular systems with memory can contribute to understanding how is "complexity" handled in nature. Indeed, memory properties are among the basic mechanisms of natural networks.[26] Near-term, new network elements with memory will result in novel computing paradigms[27] and add to the emerging "toolbox" of biocomputing and multi-signal-biosensing systems and components.[28]

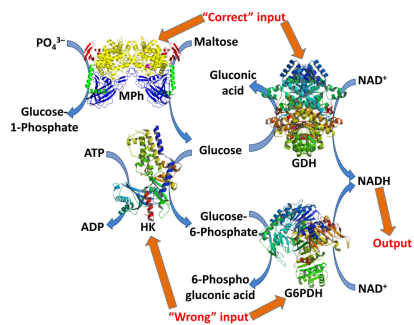

**Scheme 1.** The biocatalytic process steps of the associative memory system based on a two-pathway reaction cascade.

In this work we report a new model system activated by enzymes as signals and accomplishing associative memory. It is important to emphasize that associative memory[29] is a concept that has been studied for a long time[30] and in diverse fields,[31] but at the same time it has been a subject of active recent research[32-35] driven by conceptual interest and also by the promise of adding adaptive capabilities to information processing systems. However, despite active research in the area, very few biomolecular systems based on proteins[36,37] or DNA[38-40] mimicking associative memory *in vitro* haven been realized. Various definitions of the process are possible.[29] In the present study we realized the associative memory where application of the "correct" input signal activates the system to yield the output signal, while the "wrong" signal results in no chemical changes. A simultaneous application of the "correct" and "wrong" signals ("training" step) results in the memory effect after which the system reacts to the application of the "wrong" signal as it would react to the "correct" signal, by producing the output signal. Later the system can "forget" the "training" and stop reacting to the "wrong" signal applied alone. Here we focus on a system in which the processes of conversion of the input signal to output are entirely those biocatalyzed by enzymes in two parallel cascades, Scheme 1, allowing a simple biochemical system to mimic the Pavlov's dog[30] behavior without complex biological mechanisms. We point out that inputting signals, detecting the output, and resetting the system are steps which will typically involve physical or chemical processes interfacing with the main biochemical steps. These enable spatial and temporal ("clocking") control of chemical concentrations by transport, separation, compartmentalization, and other physical or chemical means, see details and also discussion of future challenges in the Electronic Supporting Information (ESI).

The system "machinery" (the constant part of the reacting system) was composed of $NAD^+$ (3 mM), ATP (5 mM), and maltose (22.5 mM) in 0.1 mM phosphate buffer saline, pH = 7.0. The "correct" signal was defined as the presence of two enzymes, maltose phosphorylase (MPh, E.C. 2.4.1.8) and glucose dehydrogenase (GDH, E.C. 1.1.1.47), which activate the primary pathway converting maltose to glucose and then use glucose for the reduction of $NAD^+$ to NADH, Scheme 1. The optical absorbance increase at $\lambda = 340$ nm corresponding to the formation of NADH was defined as the output signal. The "wrong" signal was defined as the presence of two enzymes, hexokinase (HK, E.C. 2.7.1.1) and glucose-6-phosphate dehydrogenase (G6PDH, E.C. 1.1.1.49), which does not result in any reaction in the system if glucose is not present in the solution. Simultaneous application of the "correct" and "wrong" signals (i.e., all four enzymes) results in the activation of both pathways, where the supply of glucose produced in the presence of MPh is split in two parallel reactions, with the secondary pathway including glucose conversion to glucose-6-phosphate (Glu6P) catalyzed by HK and then Glu6P is used for the reduction of $NAD^+$ to yield NADH. Note that the output signal is produced by both pathways, Scheme 1. The system composition was optimized in such way that the "training" step (application of the "correct" and "wrong" signals together) resulted in accumulation of the intermediate product Glu6P representing the system memory. In order to achieve this, the activity of G6PDH was 5-fold smaller than the activity of HK, thus providing the production of Glu6P faster than its consumption. Later application of the "wrong" signal alone results in the output signal (production of NADH) using Glu6P as a reductant accumulated in the "training" step. Note that the activities of all enzymes (particularly of HK and GDH) were optimized experimentally to split glucose into two pathways in such a way that the output signal is preserved almost at the same level at the reaction time. After each signal application step the enzymes were removed from the solution by ultrafiltration and $NAD^+$, ATP and maltose were added to the solution in order to have them at the same concentrations for the next signal application step.

Figure 1 shows a succession of steps involving applications of different inputs. The "correct" input produced finite output signal, whereas during the application of the "wrong" input only, no signal is produced. However, when both the "correct" and "wrong" inputs were applied together (in the "training" step) the signal was produced again through both biocatalytic cascades. As mentioned, the system was tailored to allow for a surplus of Glu6P to be produced *in situ* during the "training" step. Therefore, in a later application of the "wrong" input only, during the step termed "memory," $NAD^+$ was reduced by G6PDH, producing a signal. Subsequent additions of the "wrong" input produce no signal, because practically all of the Glu6P-based memory is used up during the first "memory" step (if all the process steps are properly timed). This demonstrates that the memory realized here is not "self-reinforcing" – the system "forgets" the training. Figure 2A demonstrates the possibility to "reset" the system after various step sequences. The biocatalytically produced NADH was recycled to $NAD^+$ by a photochemical reaction[41] in the presence of thionine and oxygen upon illumination (see the experimental details in the ESI). This was done after each step sequence to illustrate that the system can be returned to its initial state at any point.

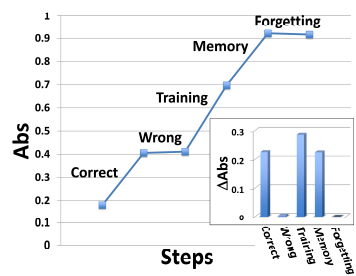

**Figure 1.** Optical measurements at 340 nm of the solution after the reaction with various input applications, as described in the text. Inset: Change in the absorbance measured after each 10 min input step, indicating the quantity of the output product.

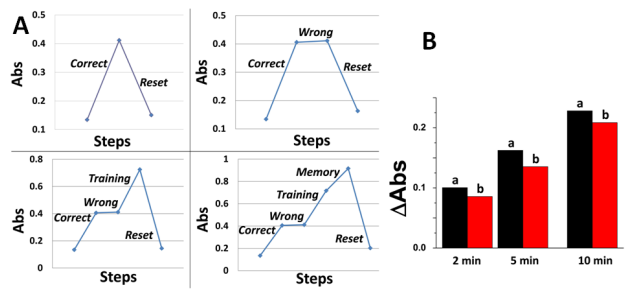

**Figure 2.** (A) Experiments demonstrating that the system can be "reset," i.e., returned to a minimum absorbance level, at any point in the succession of reactions steps. (B) Time dependence of "training" (a) and "memory" (b) for varying "training" times: 2, 5 and 10 minutes. The "memory" step was carried out for the same amount of time, 10 min, in each case.

The effect of the "training" step duration time on "memory" was also tested, as shown in Figure 2B. This was done by varying the initial incubation time of the "correct" and "wrong" inputs jointly, allowed to proceed for 2, 5, and 10 min. Measurements were then taken with the "wrong" input only, as the "memory" step for 10 min. Notably the longer "training" step results in more pronounced "memory" step due to a larger accumulation of Glu6P.

Our results illustrate rather general properties specific to biocatalytic systems used as memory processes. We expect that such systems will be utilized in multi-input and/or multi-step setups with digital, binary-signal information processing steps temporally controlled ("clocked") and spatially controlled / compartmentalized for sensor and signal/information processing applications. Each of the process steps will then be timed, and signal levels will have to be maintained at well-defined values. The present system functions under rather stringent conditions, limiting its flexibility. This is further commented on in a section on future challenges in the ESI, where we also note that it would be interesting to address the non-steady state time-dependence regimes required for accumulation and depletion of "memory" chemicals in enzymatic memory systems,[32] as well as devise flexible system and develop ideas for their networking. Networking of memory elements for useful information processing is a rather new, recently emerging topic in electronics.[42,43] An associative memory system should preferably be devised in such a way that in the "training" step, the added "wrong" input does not significantly modify the output intensity (as seen in Figure 1). Furthermore, after the "training," application of the "wrong" output alone should result in an output rate comparable to the original rate (cf. Figure 1). However, the accumulated chemical will eventually be exhausted and the system will "forget" the training, and after one or more fixed-duration time steps the product will no longer be generated. Based on the experiments shown in Figures 1-2 we conclude that our system shows all the aspects of this expected pattern of behavior, including fast "forgetting" practically after a single "memory" step.

In summary, we reported the first demonstration of the associative memory as a bio-inspired memory-involving signal processing system functioning with the internal dynamics entirely biocatalyzed by enzymes, activated by enzymes as inputs. This model system shows general characteristics expected of memory "devices" based on chemical compound accumulation. It is hoped that such systems can offer new "network elements" and novel bio-inspired design ideas for biomolecular logic being developed for biosensors[11] and biocomputing[14] applications.

This research was supported by the National Science Foundation (awards CBET-1066397 and CCF-1015983). The authors thank Nora Kohlstedt for the drawing of the Pavlov's dog in the graphical abstract.

## Notes and references

*a* Department of Chemistry & Biomolecular Science, *b* Department of Physics, Clarkson University, Potsdam, NY 13699, USA. Fax: 1-315-268-6610; Tel: 1-315-268-4421; E-mail: ekatz@clarkson.edu
† Electronic Supplementary Information (ESI) available: Detailed experimental procedures, and discussion of future challenges. See DOI: 10.1039/C3CC43272F

# Bioinspired Associative Memory System Based on Enzymatic Cascades

Kevin MacVittie,[a] Jan Halámek,[a] Vladimir Privman,[b] and Evgeny Katz[a]*

[a] Department of Chemistry & Biomolecular Science, [b] Department of Physics, Clarkson University, Potsdam, NY 13699, USA

## Electronic Supplementary Information (ESI)

*Chemicals and Materials.* The following enzymes for the biochemical associative memory system were obtained from Sigma-Aldrich and used without further purification: maltose phosphorylase (MPh) from *Enterococcus sp.* (E.C. 2.4.1.8), glucose dehydrogenase (GDH) from *Pseudomonas sp.* (E.C. 1.1.1.47), hexokinase (HK) from *Saccharomyces cerevisiae* (E.C. 2.7.1.1), and glucose-6-phosphate dehydrogenase (G6PDH) from *Leuconostoc mesenteroides* (E.C. 1.1.1.49). Other chemicals from Sigma-Aldrich used as supplied included: D-(+)-maltose monohydrate, adenosine 5'-triphosphate disodium salt hydrate (ATP), and β-nicotinamide adenine dinucleotide sodium salt ($NAD^+$). Thionine acetate was obtained from Alfa Aesar. Ultrapure water (18.2 MΩ cm) from NANOpure Diamond (Barnstead) source was used in all of the experiments.

*Instrumentation and Measurements.* A Shimadzu UV-2401PC/2501PC UV-Vis spectrophotometer (Shimadzu, Tokyo, Japan) with 1 mL poly(methyl methacrylate) (PMMA) cuvettes was used for all measurements. The halogen bulb lamp (500 W T3 halogen bulb) was purchased from ACE Hardware Corp. During the photochemical reset step, the reaction solution was illuminated with non-filtered polychromatic light from the lamp with the intensity measured with Light Meter LX802 (MN Measurement Instruments).

*The system composition and reaction steps.* The core "machinery" of the biochemical associative memory system was composed of $NAD^+$ (3 mM), ATP (5 mM), and maltose (22.5 mM) in 0.1 mM phosphate buffer saline (PBS), pH = 7.0 PBS included 137 mM NaCl, 2.7 mM KCl, 10 mM $Na_2HPO_4$ and 2 mM $KH_2PO_4$ titrated to pH = 7.0. MPh (1 U $mL^{-1}$) and GDH (2 U $mL^{-1}$) were used as the "correct" input. The "wrong" input consisted of HK (10 U $mL^{-1}$) and G6PDH (2 U $mL^{-1}$). Reactions were carried out in 3 kDa micro-centrifuge tubes (Pall Corporation, Michigan) at ambient temperature 23 °C. Reactions after applying the signals ("correct", "wrong", "correct"+"wrong") in various combinations were allowed to proceed for 10 min (unless stated differently) before the solution was removed by ultrafiltration: by centrifugation at 12,000 rpm in a Microfuge 22R centrifuge (Beckman Coulter, CA). After separation of the supernatant from the used enzyme inputs, the "machinery" components ($NAD^+$, ATP and maltose) were added to the solution and then the next enzyme input was applied. The "reset" operation returning the NADH output back to $NAD^+$ (thus resetting the optical absorbance at $\lambda = 340$ nm to its initial value) was performed by the irradiation of the solution with the non-filtered light from the halogen lamp with intensity of 9,000 lux for 5 min in the presence of thionine (250 μM) and oxygen (in equilibrium with air).

*Comments on Future Challenges.* The present realization of a purely enzymatic associative memory is limited to a rather specific set of conditions. Future work should focus on ideas of improving flexibility and versatility of enzymatic and other bio-inspired memory elements for enabling their networking. As referenced in the main text, networking concepts for useful information processing are rather recent even in electronics, and will have to be developed for biomolecular systems. The "toolbox" of future gates should also include some modularity, i.e., flexibility in which compounds to use and what is their role. For example, in our system the accumulating compound is Glu6P. Changing this, for instance to have glucose accumulate instead, for the memory effect in an otherwise similar system would require complete redesign and use of different operating regimes. The time-evolution of memory systems realized as purely enzymatic biocatalytic cascades should also be studied for kinetic regimes involved. In typical kinetic studies of enzymatic reactions with the system kept under steady state conditions for the time intervals of the experiment. In memory systems, however, accumulation/depletion of intermediate chemical(s) suggests non-steady-state kinetics for both the "learning" and "forgetting" processes.

Table of Contents Entry

## Associative memory: complex and simple

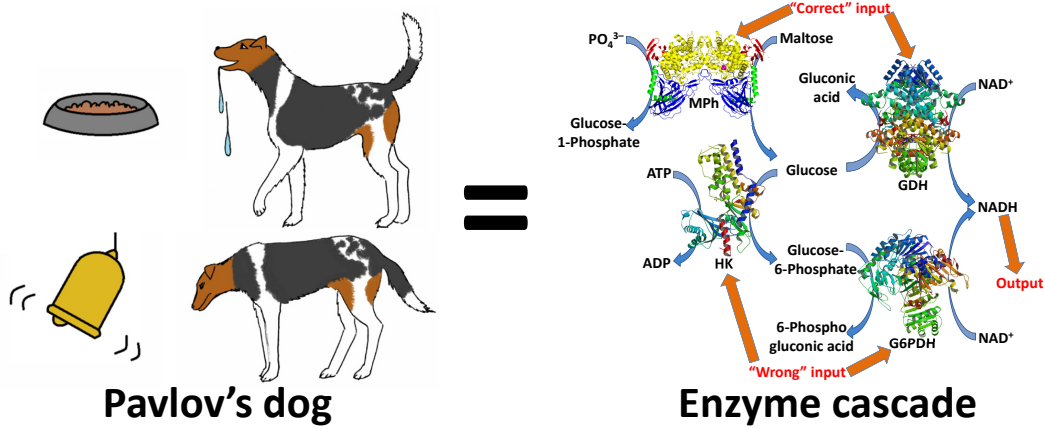

**Pavlov's dog** = **Enzyme cascade**

A biomolecular system representing the first realization of associative memory based on enzymatic reactions *in vitro* has been designed. The system demonstrated "training" and "forgetting" features characteristic of memory in biological systems, but presently realized in simple biocatalytic cascades.